\documentclass[10pt]{article}
\usepackage[utf8]{inputenc}
\usepackage{authblk}
\usepackage{silence}
\usepackage{tcolorbox} 
\usepackage[a4paper, left=1in, right=1in, top=1in, bottom=1in]{geometry}
\usepackage{setspace}
\usepackage[symbol]{footmisc}
\usepackage[hidelinks]{hyperref}
\newcommand{\quotes}[1]{``#1''}
\title{Predictive processing frameworks for perception can explain recent drone sightings in the United States}
\author[1,2]{Joel Frohlich}
\author[1]{Leonardo Christov-Moore}
\author[1]{Nicco Reggente}

\affil[1]{Institute for Advanced Consciousness Studies, Santa Monica, California, USA}
\affil[2]{fMEG Center, University of Tuebingen, Tuebingen, Germany}

\begin{document}

\onehalfspacing

\maketitle
Late 2024 was marked by mass delusion on a scale echoing the notorious 1938 radio broadcast of \quotes{War of the Worlds}. Beginning on 18 November, over 5,000 reports of mysterious drones were submitted to the FBI across New Jersey, with similar sightings quickly spreading to other parts of the United States \cite{DHSFBIJoint2024} . Now documented as the \quotes{2024 United States drone sightings}, the incident(s) accumulated thousands of witnesses, some of whom perceived hovering objects \quotes{the size of bicycles or small cars} that reportedly operated in coordinated patterns and could turn off their lights to evade detection \cite{CNNNJDS2024} . The incident influenced mainstream political discourse, with President-elect Donald Trump posting on Truth Social that authorities should \quotes{Let the public know, and now. Otherwise, shoot them down!!!} \cite{SalonTrump2024} . While these objects were frequently discussed in a geopolitical context—including claims of Iranian \quotes{mothership} deployment, promptly refuted by the Pentagon \cite{BBCDrones2024} — the cultural undercurrents driving such reports echo those of UFO sightings. 

Many witnesses proved incorrigible despite federal agencies concluding many sightings were misidentified aircraft --- indeed, some witnesses even directed laser pointers at commercial flights \cite{sweeney2024laser, hoel2024mass}. This steadfast adherence to misperception is particularly striking given that these lights in the sky are rather easily explained through conventional means. Nonetheless, the widespread and persistent misperception of planets, stars, and commercial passenger jets as mysterious drones by so many people seems baffling.  While social media, confirmation bias, and mass \quotes{hysteria} have been put forward as partial explanations \cite{Hoel2024}, we believe that more fundamental principles of conscious perception --- mainly, predictive processing and Bayesian inference --- offer possible mechanisms. 

Here, we make a unique contribution to the discussion by drawing on the predictive processing theory of perception (Box 1) to explain why \textbf{healthy, intelligent, honest, and psychologically normal people might easily misperceive lights in the sky as threatening or extraordinary objects, }especially in the context of WEIRD (western, educated, industrial, rich, and democratic) societies. We argue that the uniquely sparse properties of skyborne and celestial stimuli make it difficult for an observer to update prior beliefs, which can be easily fit to observed lights. We briefly discuss the possible role of generalized distrust in scientific institutions \cite{Achterberg2017} and ultimately argue for the importance of astronomy education for producing a society with prior beliefs that support veridical perception. 

\begin{tcolorbox}[title=\textbf{Box 1: Predictive processing}, colback=blue!5, colframe=blue!80!black, sharp corners=southwest]
According to predictive processing (also known as \quotes{predictive coding}), our conscious experiences best correspond to an internal, prior model of the world generated by the brain \cite{Bottemanne2025}. This prior model is compared with sensory input and updated to minimize prediction errors. Because this process follows principles of Bayesian inference, predictive processing belongs to a broader collection of theories often referred to as the \quotes{Bayesian brain hypothesis}, which posits that \quotes{we are trying to infer the causes of our sensations based on a generative model of the world} \cite{friston2012history}.
\end{tcolorbox}

\section*{The sky is uniquely challenging for perception}

Our first insight toward understanding the drone mass-sighting incident, as well as the broader phenomena of UFO and UAP (unidentified aerial phenomena) sightings, is that \textit{the sky and objects that appear against it have unique stimulus properties}. Henceforth, we refer to this as the\textbf{ Principle of Skyborne Impoverishment (PSI)}. This principle has three important dimensions; unlike objects that appear on land, objects which appear in the sky \textbf{1) lack context}, as the sky often appears as a homogeneous field, \textbf{2) span many orders of magnitude in distance}, from meters to light years, and \textbf{3) appear as approximately zero-dimensional points}, precluding perceptual convergence based on shape, texture, or other stimulus properties that are useful at closer range. This represents a particularly exacerbated instance of the \quotes{Inverse Optics Problem} in vision science \cite{Zygmunt2001} as well as a breakdown of  \quotes{size constancy} mechanisms-- our visual system's ability to maintain stable perception of object size despite variations in viewing distance \cite{Sperandio2015}.

Due to the PSI, one's perception of lights that appear in the sky is largely top-down or prior heavy, i.e., strongly driven by one's current generative model of the world. This is because visual afferent signals corresponding to a point-like light in the sky can be easily fit to many prior models without generating a substantial prediction error. The same visual signals might fit with a drone, an approaching airplane, the planet Jupiter, a star, or even a triple star system; see Figure 1.

\begin{figure*}[ht]
\centering
\includegraphics[width=\linewidth]{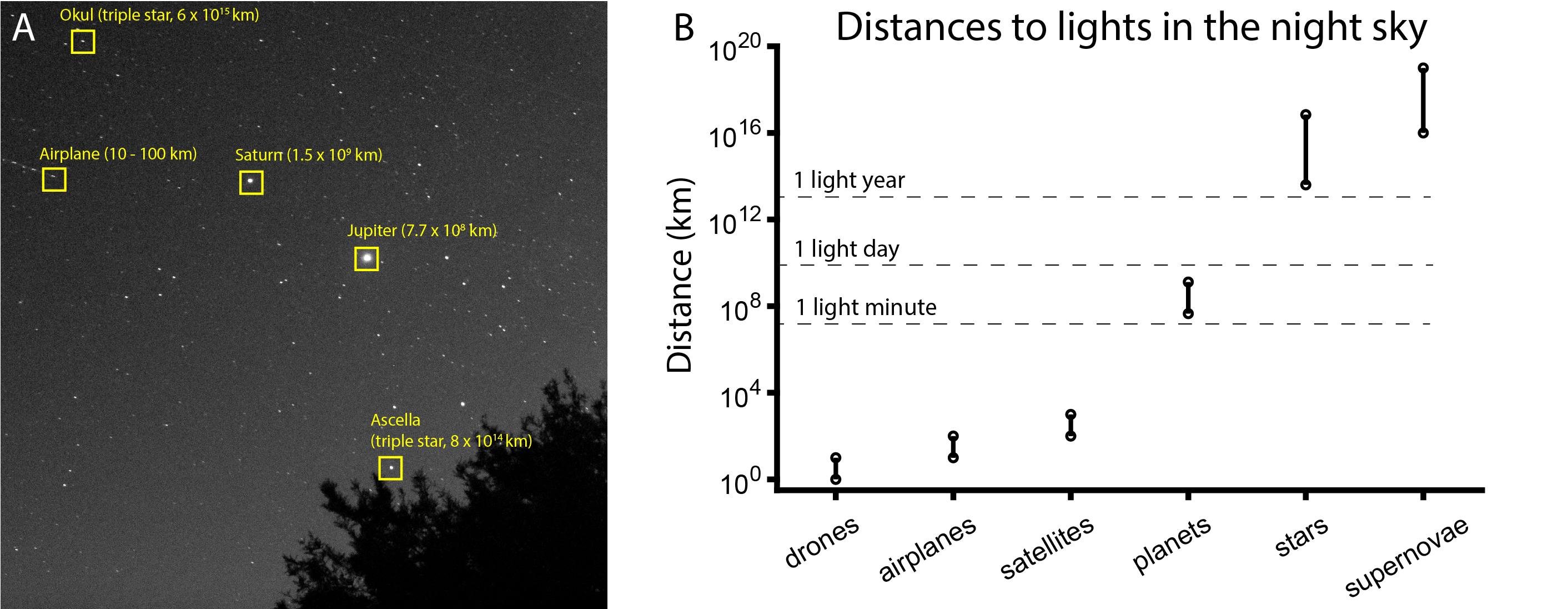}
\caption{\textbf{The principle of skyborne impoverishment (PSI).} As viewed by eye from the ground, objects spanning many orders of magnitude in distance may appear as discrete points that are easily fit to different prior models without generating a prediction error. The photograph in panel A is an annotated 24 s exposure taken by author JF facing southwest from Joshua Tree, CA, USA on 19 October 2020 with a Canon 70D camera (f/5.6, iso 6400). While this photographic image demonstrates the PSI, note that the camera cannot capture exactly the same perspective as the human visual system, e.g., due to the long exposure, the plane's blinking navigation lights appear as three separate points, and other objects are slightly elongated due to the Earth's rotation. Panel B illustrates the distances of objects that may appear as pointlike lights to the unaided eye on a logarthmic distance scale.}
\label{fig:Fig1}
\end{figure*}


\section*{Humans view the sky as inherently meaningful}

Our pre-industrial ancestors looked to the sky to understand time, seasons, weather, and navigation, knowledge of which carries, in each case, strong survival value. As a result, humans have likely evolved to perceive the sky as deeply meaningful. Evolution by natural selection unfolds on a slow timescale relative to technological and societal progress. As a result, we continue to view the sky, rather than our computers or smartphones, as deeply meaningful. The sky offers a rare perceptual paradox: it is wholly visible yet utterly untouchable—an interface between the tangible and the ineffable, the known and the unknowable. Unlike the microscopic world of cells and molecules, which we can only see, at best, with special instruments, the sky reveals itself freely to the naked eye, yet resists grasp. And it is often at such boundaries—where observation collides with mystery, where we can see but not fully understand—that we are most compelled to assign meaning. We witness babies born yet cannot fully comprehend how life emerges from strands of DNA; we look to the stars with the same awe.  This hypothesis may explain cross-cultural associations of the sky with heaven or an afterlife, as well as the longevity of belief in astrology, an ancient pseudoscience which puts forth the apparent positions of celestial objects to explain human affairs and, remarkably, has flourished in different forms across many cultures \cite{britannica_astrology}.

In a nutshell, we hypothesize that \textit{the sky is psychedelic}. The psychiatrist Humphry Osmond originally coined the term \quotes{psychedelic} in 1956 to explain the ability of certain psychoactive drugs, such as LSD (lysergic acid diethylamide) and psilocybin, to \quotes{manifest the mind} or reveal subconscious beliefs and emotions \cite{mcintyre2023psychedelic}. Scientific research has shown that LSD and psilocybin alter perception, in part, by revealing profound meaning in otherwise ordinary situations, persons, and objects. They push perception to that same threshold-- where the seen brushes against the unknowable, and what is familiar becomes freshly mysterious. Similarly, our \textbf{Psychedelic Sky Hypothesis (PSH)} posits that the sky is also a \quotes{meaning amplifier} which, like a psychedelic drug, manifests subconscious beliefs and emotions. As with psychedelic drugs, these beliefs and emotions may be deeply positive --- the joy of perceiving extraterrestrial visitors that sometimes accompanies UFO sightings \cite{hernandez2018study} --- or deeply negative --- the fear that may accompany drone sightings attributed to a nefarious foreign adversary. 

\section*{Industrial societies view the sky with the wrong priors}

According to the PSH, humans have perceived the sky as deeply meaningful for millennia. Why, then, are UFO and drone sightings peculiar to modernity and, arguably, WEIRD cultures \cite{NUFORC_map, nugent2017ufo} (see Figure 2)? An obvious consideration is that preindustrial people had no concept of drones, much less of spacecraft (c.f., \cite{stothers2007ufo} ). Less obviously, however, many preindustrial people likely enjoyed (and continue, in the developing world, to enjoy) a level of astronomical literacy that is rare in today's industrial societies. Industrial, urbanized societies experience the night sky very differently from their preindustrial ancestors due to diffuse light from artificial sources. Whereas preindustrial people could view the Milky Way galaxy and thousands of stars with the unaided eye, industrial city dwellers can view only a fraction of these stars through the veil of light pollution. This, coupled with the ubiquity of smartphone devices, may also produce a scenario in which individuals simply attend less often to the sky overall.

\begin{figure*}[ht]
\centering
\includegraphics[width=\linewidth]{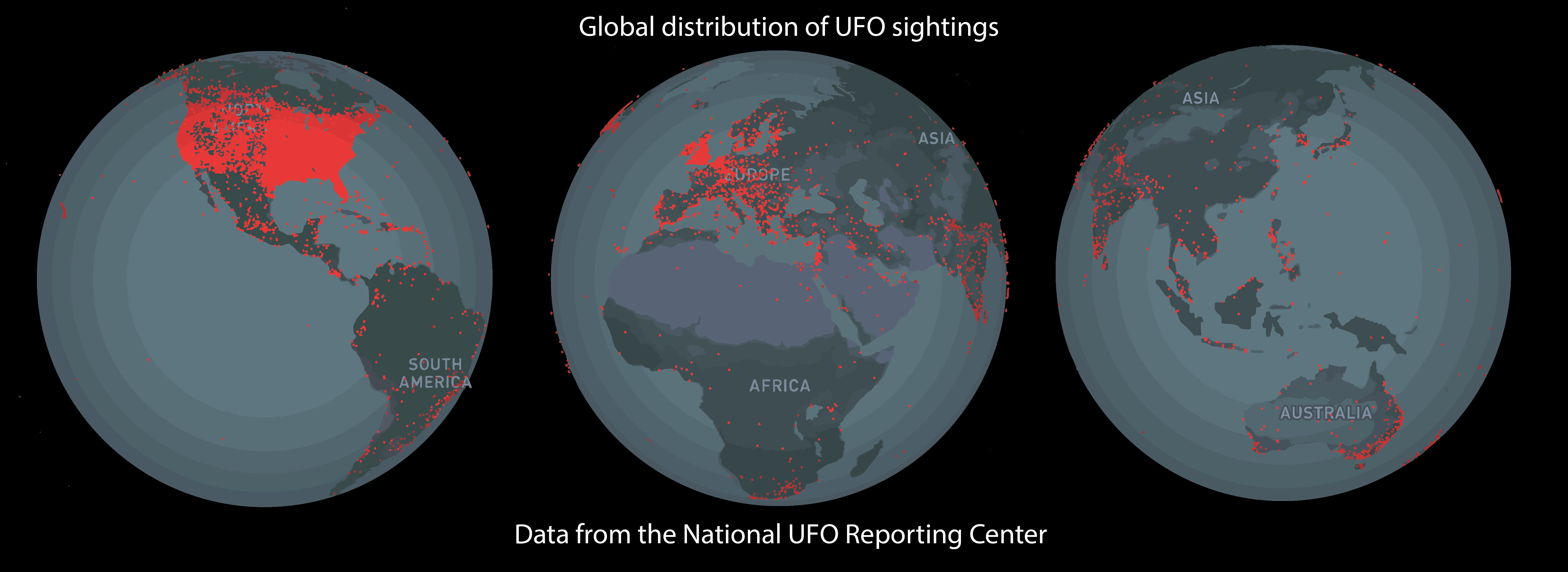}
\caption{\textbf{UFO sightings are common in WEIRD countries.} Map images were adapted from an online map of UFO sightings compiled by the National UFO Reporting Center (Davenport, WA, USA) database (n = 154486 sightings) and used with permission. UFO sightings locations are indicated in red. Even given the possibility of a sampling bias, the data suggest that UFOs are reported more often in the industrial Western world, with the greatest number of reports coming from the USA, UK, Canada, and Australia. Of 154486 UFO reports, \(< 0.1\%\) were dated before 1950.} 
\label{fig:Fig2}
\end{figure*}

It is often argued that astronomy is the oldest science; even in antiquity, many preindustrial people would have incorporated the planets into their generative world model, going at least as far back as ancient Mesopotamia. By contrast, because astronomy education does not have obvious economic or vocational value in the industrial world, many otherwise well-educated people in WEIRD societies are ignorant of the fact that five other planets in our solar system are visible to the unaided eye, often outshining stars. For instance, the planets Jupiter and Venus often appear several times brighter than the brightest star in the night sky, Sirius\footnote{Sirius: magnitude -1.47 Jupiter: max magnitude -2.7, Venus: max magnitude -4.4; each magnitude is 2.51 times brighter than the previous (fifth root of 100). \cite{Dutkevitch1998}}. This fact, coupled with the more stable appearance of planets relative to \quotes{twinkling} stars \cite{whitehead2012stars}, gives the planets properties as stimuli that are noticeably not \quotes{star-like}. When a person in today's industrial world fails to recognize a planet as such, their mind may instead converge on the percept of a drone.

The above issue is further complicated by artificial satellites and space stations. The number, brightness, and configuration of satellites have increased dramatically in the past several decades and, with the advent of the Starlink satellite constellation, even in the past several years. A Starlink satellite train appears as dozens of points of light moving in a straight line across the night sky, an unprecedented occurrence often perceived as a UFO formation. And since its completion in 2011, the International Space Station is occasionally the brightest object in the night sky after the Moon. Much of the general public may remain unaware that satellites and space stations are often visible to the unaided eye, even within the industrial societies that constructed them.  

\section*{The free energy principle may explain the incorrigibility of witnesses}

Even if one's perception of a drone or UFO is later challenged by reading contrary evidence from experts, this evidence may be insufficient for updating one's prior model of reality, for it is \textit{the wrong kind of evidence}. According to predictive processing, the brain aims to minimize the discrepancy between a prior model of the world and afferent sensory signals which convey information about the world (Box 1). In one framework for understanding predictive processing, updates to internal models are primarily driven by mismatches between predictions and direct sensory inputs, rather than discrepancies stemming from second-hand reports or abstract knowledge (e.g., information conveyed by a skeptical expert). A related framework, known as the Free Energy Principle (FEP), hypothesizes that all biological organisms have a fundamental drive to decrease variational  \quotes{free energy}-- a measure that bounds surprise by quantifying the divergence between internal predictions and sensory states. 

While in principle any source of information can reduce free energy, the brain assigns varying levels of precision—or confidence—to different inputs. Direct sensory signals, given their immediate relevance for survival in our evolutionary past, are typically weighted more heavily in the minimization of free energy. In contrast, expert analysis is often conveyed linguistically and after the perceptual event, making it less likely to influence the generative model responsible for shaping the original experience. Such information seldom coincides with the sensory data it seeks to reinterpret, and thus may fail to update the relevant perceptual priors.

However, as recent work suggests, social cues—particularly from trusted officials—can act as higher-order priors themselves, modulating how incoming information is interpreted and integrated \cite{kaplan2023moral}. In this view, trust functions as a kind of epistemic gatekeeper \cite{christov2024toward}, increasing the precision weighting of certain abstract inputs and thereby granting them greater influence over belief updating. Yet even these socially conferred priors may be insufficient to revise deeply entrenched perceptual expectations if they do not significantly alter predicted sensory consequences. As a result, while expert opinions may still generate cognitive dissonance, this dissonance is often resolved through higher-level mechanisms like confirmation bias or motivated reasoning \cite{stone2018cognitive}, rather than through perceptual systems that govern how the event was originally seen—or remembered.

\begin{figure*}[ht]
\centering
\includegraphics[width=\linewidth]{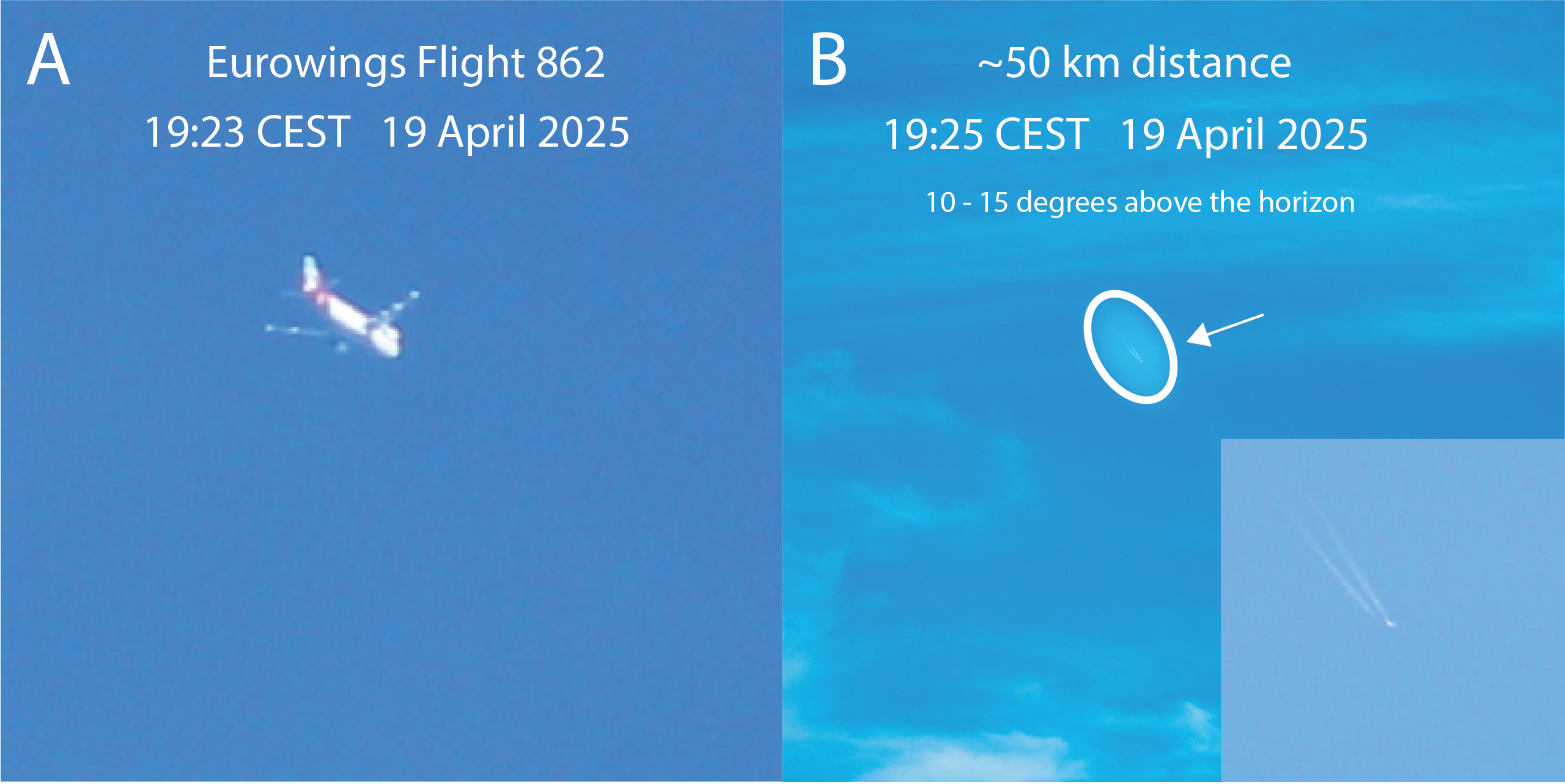}
\caption{\textbf{An airplane cruising at an altitude of 11,750 meters appears in the sky at approximately 50 km distance between 10 and 15 degrees above the horizon.} Both photographs were taken by author JF from his home in Tuebingen, Germany using a Canon 70D camera (f/5.6, iso 200, 2 ms exposure). The identity and distance of the airplane was determined using the smartphone app Plane Finder (Pinkfroot Limited). While a witness might describe seeing the airplane above their town, the distance is comparable to that between downtown Los Angeles and Irvine, California. As a consequence of the PSI, witnesses may then fail to correctly identify an airplane even with the aid of a flightracker app, given the intuition that the airplane is much closer. At night, navigation lights from airplanes may be visible from even greater distances, e.g., 100 km.} 
\label{fig:Fig3}
\end{figure*}

\section*{Conclusions}

Herein, we have argued that even intelligent and generally well-educated individuals may be susceptible to the non-veridical perception of celestial objects, satellites, and airplanes as drones or, more generally, UFOs. Furthermore, in the context of predictive processing, we have argued that the PSI results in little if any prediction error signal, impeding the updating of prior beliefs that occurs more commonly for information-rich terrestrial stimuli. Finally, the PSH predicts that the ambiguity of skyborne stimuli imposed by the PSI will often be resolved by converging toward a deeply meaningful percept. 

While the specific timing of recent drone sightings may be attributable to a combination of factors --- social media, political anxiety, institutional distrust, and the proliferation of smartphone cameras that record and even further impoverish sparse stimuli\footnote{Smartphone cameras have poor sensitivity to relatively dim lights recorded at night and compress video and images, resulting in further impoverishment and, often, compression artifacts.} --- we believe that the principles outlined above largely explain the extraordinary number of recent drone sightings in late 2024. This wave of drone sightings, and the broader phenomenon of UFO or \quotes{UAP} sightings that preceded it, has had real consequences for public policy, government spending, and political discourse, e.g., the creation of the US Department of Defense All-domain Anomaly Resolution Office (AARO) \cite{kirkpatrick2024here, hoel2024mass}. Thus, we advocate for educational reforms. Modern education should produce a society with better priors for viewing the sky, and better tools to independently evaluate the claims of scientific institutions. Specifically, valuing astronomy literacy and education in K-12 education would allow a new generation of industrialized society that is better able to perceive and understand the sky and its contents. 

While this editorial has largely focused on the misperception of celestial objects, as previously noted, airplanes are also frequently mistaken for drones or UFOs. It is likely that many people underestimate the altitude and distance of airplanes \cite{west2012contrail} (Figure 3), resulting in priors that make the misidentification of aircraft easy; for instance, a witness may not realize that the source of the light they are viewing might be over 100 km away, in the case of a distant airplane at cruising altitude \cite{west2012contrail}; thus, they may fail to identify an airplane as the light source even with the aid of a flight-tracker smartphone app. A further challenge for K-12 education will be to emphasize aviation in education and build better intuitions for observing aircraft. 

Finally, we anticipate criticism from those who may feel we are too dismissive of drone witnesses. Drones, i.e, uncrewed aerial vehicles, are indeed real. Hobbyist drones have become common in the past decade, and a high-altitude balloon originating from China was indeed discovered spying on the United States in 2023. If drones were indeed responsible for the mass sightings of 2024, we welcome all such evidence that would falsify the account put forth herein. More generally, we also welcome any other data that might falsify the PSI and PSH. For instance, if similar unusual sightings or misperceptions were shown to be common in submerged ocean environments were the sky is not visible, then our current account would be incomplete. 

As of today, thousands of drone reports in late 2024 have turned up very few confirmed drones. Even if a small number of reports were actual sightings of hobbyist drones, the vast majority are needing of another explanation. While many commentators have focused on which \textit{objects }might explain these sightings, we hope our editorial will advanced the discussion on which \textit{perceptual mechanism}s might explain these sightings. Far from intending to belittle witnesses, we believe that these individuals are often smart, educated, and honest people reporting their genuine --- albeit non-veridical --- experiences. 

\section*{Acknowledgements}

We are grateful to the National UFO Reporting Center for making available the map data used in Figure 2.

\bibliography{refs}

\end{document}